\documentclass[showpacs,aps,prd]{revtex4}
\input epsf

\begin{document}

\title{Looking for a vector charmonium-like state $Y$ in $e^{+}e^{-}\rightarrow\bar{D}D_{1}(2420)+c.c.$}
\author{Jian-Rong Zhang}
\affiliation{Department of Physics, College of Liberal Arts and Sciences, National University of Defense Technology,
Changsha 410073, Hunan, People's Republic of China}

%\date{\today}

%%%%%%%%%%%%%%%%%%%%%%%%%%%%%%%%%%%%%%%%%%%%%%%%%%%%%%%%%%%%%%%%%%%%%
\begin{abstract}
Inspired by the first observation of a vector
charmonium-like state $Y(4626)$ decaying to
a meson pair
$D_{s}^{+}D_{s1}(2536)^{-}$,
which could be viewed as a $P$-wave scalar-scalar
$[cs][\bar{c}\bar{s}]$ tetraquark state,
we predict a potential vector
charmonium-like state $Y$ with $P$-wave scalar-scalar
$[cq][\bar{c}\bar{q}]$ configuration.
The corresponding mass spectrum
of $Y$ state is calculated to be
$4.33^{+0.16}_{-0.23}~\mbox{GeV}$
in the framework of QCD sum rules.
We suggest that
the predicted $Y$ state
could be looked for in an open-charm $e^{+}e^{-}\rightarrow\bar{D}D_{1}(2420)+c.c.$ process.
\end{abstract}
\maketitle

\section{Introduction}\label{sec1}

In recent years, a series of vector charmonium-like $Y$ states
have been observed in the initial-state radiation processes $e^{+}e^{-}\rightarrow \gamma_{ISR}\pi^{+}\pi^{-}J/\psi$ ($\psi(2S)$)
\cite{Y-B,Y4360-1,Y4360-2,Y-B1,Y-B2,Y-B3,Y4360-3,Y4360-4}
or in the direct processes $e^{+}e^{-}\rightarrow\pi^{+}\pi^{-}J/\psi$ ($\psi(2S)$) \cite{Y-CLEO,Y-BESIII-1,Y-BESIII-2,Y-BESIII-3}. These experiments show
that $Y$ states mainly couple to hidden-charm
final states. In contrast, Belle newly reported the first observation of $Y(4626)$
in an open-charm process $e^{+}e^{-}\rightarrow D_{s}^{+}D_{s1}(2536)^{-}+c.c.$ with a significance of $5.9\sigma$
\cite{Y4626}, which has promptly
attracted much attention \cite{Y4626-1,Y4626-2,Th-charmonium,Th-molecular,Th-molecular-1,Th-molecular-2,Th-tetraquark,Th-tetraquark-1,Th-tetraquark-2,Th-Zhang}.
Theoretically, some authors pointed that $Y(4626)$ can be well interpreted as
a $P$-wave $[cs][\bar{c}\bar{s}]$ state
with a multiquark
color flux-tube model \cite{Th-tetraquark-1}. Moreover, we studied $Y(4626)$
from two-point QCD sum rules,
and finally arrived at that it could be a $P$-wave scalar-scalar $[cs][\bar{c}\bar{s}]$ state \cite{Th-Zhang}.
On analogy of $Y(4626)$'s observation in the open-charm process,
we propose that
a novel vector charmonium-like state $Y$
could be looked for in an open-charm $e^{+}e^{-}\rightarrow\bar{D}D_{1}(2420)+c.c.$ process.
In theory, the predicted $Y$ state
could correspondingly be regarded as a $P$-wave scalar-scalar $[cq][\bar{c}\bar{q}]$ tetraquark state.

In this work, we endeavor to explore the
charmonium-like state $Y$ with $P$-wave scalar-scalar
$[cq][\bar{c}\bar{q}]$ configuration.
To deal with the hadronic state, one has to
confront the complicated nonperturbative QCD problem.
As one trusty method for evaluating nonperturbative effects,
the QCD sum rule \cite{svzsum} is firmly founded on the basic QCD theory,
and has been successfully applied to plenty of hadronic systems (for reviews see Refs.
\cite{overview,overview1,overview2,overview3} and references
therein).
Accordingly, we intend to study
this $Y$ state by making use of the
QCD sum rule approach.

The paper's organization is as follows. In Sec. \ref{sec2}, the QCD sum rule
is derived for $Y$ with $P$-wave scalar-scalar $[cq][\bar{c}\bar{q}]$ structure,
along with numerical analysis and discussions in Sec.
\ref{sec3}. The last part includes a brief summary.

%%%%%%%%%%%%%%%%%%%%%%%%%%%%%%%%%%%%%%%%%%%%%%%%%%%%%%%%%%%%%%%%%%%
\section{the QCD sum rule for $Y$ with $P$-wave scalar-scalar $[cq][\bar{c}\bar{q}]$ structure}\label{sec2}

Generally speaking, one could have several choices on diquarks
to characterize a $P$-wave tetraquark state with $J^{P}=1^{-}$.
It is worth noting that there have been broad discussions on the so-called
``good" or ``bad"
diquarks for the tetraquark states \cite{diquarks},
and then the $Y$ state with $P$-wave $[cq][\bar{c}\bar{q}]$ structure
could be represented basing on following considerations \cite{current}.
A ``good" diquark operator in the attractive anti-triplet color channel
can be $\bar{q}_{c}\gamma_{5}q$ with $0^{+}$,
and a ``bad" diquark operator can be $\bar{q}_{c}\gamma q$ with $1^{+}$.
Similarly, operators with $0^{-}$ and $1^{-}$ can be written as $\bar{q}_{c}q$
and $\bar{q}_{c}\gamma\gamma_{5}q$, respectively. Further,
it is suggested that diquarks are preferably formed
into spin $0$ from lattice results \cite{Lattice}.
Comparatively, the solid tetraquark candidates tend to be composed of $0^{+}$ ``good" diquarks.
For example,
the final results from QCD sum rules favor the scalar diquark-scalar antidiquark case
after comparing different diquark configurations \cite{scalar-scalar}.
Thereby, the predicted $Y$ state would be dominantly structured
as the $P$-wave scalar diquark-scalar antidiquark,
which contains the flavor content $[cq][\bar{c}\bar{q}]$ with momentum
numbers $S_{[cq]}=0$, $S_{[\bar{c}\bar{q}]}=0$, $S_{[cq][\bar{c}\bar{q}]}=0$, and $L_{[cq][\bar{c}\bar{q}]}=1$.
Here $q$ can be $u$ or $d$ quark, and
$c$ is the charm quark. Considering that both light $u$ and $d$ quark masses are taken as
current-quark masses in the paper,
they are so small comparing with
the heavy running charm mass $m_{c}$
that they will be neglected in the calculation
complying with the usual treatment of heavy hadrons.
Thus it is not concretely differentiated
whether $q=u$ or $q=d$
for brevity.
The corresponding current could be constructed as
\begin{eqnarray}
j_{\mu}=\epsilon_{def}\epsilon_{d'e'f}(q_{d}^{T}C\gamma_{5}c_{e})D_{\mu}(\bar{q}_{d'}\gamma_{5}C\bar{c}_{e'}^{T}),
\end{eqnarray}
in which the index $T$ denotes matrix
transposition, $C$ means the charge conjugation matrix,
$D_{\mu}$ is the covariant derivative to generate $L=1$, and $d$, $e$,
$f$, $d'$, and $e'$ are color indices.

Generally, the two-point correlator
$\Pi_{\mu\nu}(q^{2})=i\int
d^{4}x\mbox{e}^{iq.x}\langle0|T[j_{\mu}(x)j_{\nu}^{+}(0)]|0\rangle$
can be parameterized as
\begin{eqnarray}
\Pi_{\mu\nu}(q^{2})=\frac{q_{\mu}q_{\nu}}{q^{2}}\Pi^{(0)}(q^{2})+(\frac{q_{\mu}q_{\nu}}{q^{2}}-g_{\mu\nu})\Pi^{(1)}(q^{2}).
\end{eqnarray}
To yield the sum rule, the part $\Pi^{(1)}(q^{2})$
can be evaluated in two different ways.
At the hadronic level,
it can be expressed as
\begin{eqnarray}\label{Ph}
\Pi^{(1)}(q^{2})=\frac{\lambda^{2}}{M_{H}^{2}-q^{2}}+\frac{1}{\pi}\int_{s_{0}}
^{\infty}ds\frac{\mbox{Im}\Pi^{(1)}(s)}{s-q^{2}},
\end{eqnarray}
where $\lambda$
is the hadronic coupling constant and $M_{H}$ is the hadron's mass.
At the quark level, it can be written as
\begin{eqnarray}\label{OPE}
\Pi^{(1)}(q^{2})=\int_{4m_{c}^{2}}^{\infty}ds\frac{\rho(s)}{s-q^{2}},
\end{eqnarray}
for which the spectral density
$\rho(s)=\frac{1}{\pi}\mbox{Im}\Pi^{(1)}(s)$.

In deriving $\rho(s)$, one could work at leading order in $\alpha_{s}$
and consider condensates up to dimension $8$.
To keep the heavy-quark mass finite, one uses
the heavy-quark propagator in
momentum space \cite{reinders}.
The correlator's light-quark part
is calculated in the
coordinate space and Fourier-transformed to the $D$ dimension momentum
space, which is combined
with the heavy-quark part and then dimensionally regularized at
$D=4$ \cite{overview3,Nielsen,Zhang}.
It is given by
$\rho(s)=\rho^{\mbox{pert}}+\rho^{\langle\bar{q}q\rangle}+\rho^{\langle
g^{2}G^{2}\rangle}+
\rho^{\langle g\bar{q}\sigma\cdot G q\rangle}+\rho^{\langle\bar{q}q\rangle^{2}}+\rho^{\langle g^{3}G^{3}\rangle}+\rho^{\langle\bar{q}q\rangle\langle g^{2}G^{2}\rangle}+\rho^{\langle\bar{q}q\rangle\langle g\bar{q}\sigma\cdot G q\rangle}$,
detailedly with
\begin{eqnarray}
\rho^{\mbox{pert}}&=&-\frac{1}{3\cdot5\cdot2^{11}\pi^{6}}\int_{\alpha_{min}}^{\alpha_{max}}\frac{d\alpha}{\alpha^{4}}\int_{\beta_{min}}^{1-\alpha}\frac{d\beta}{\beta^{4}}(1-\alpha-\beta)\kappa
r^{5},\nonumber\\
\rho^{\langle\bar{q}q\rangle}&=&\frac{m_{c}\langle\bar{q}q\rangle}{3\cdot2^{6}\pi^{4}}\int_{\alpha_{min}}^{\alpha_{max}}\frac{d\alpha}{\alpha^{2}}\int_{\beta_{min}}^{1-\alpha}\frac{d\beta}{\beta^{2}}(2-\alpha-\beta)r^{3},\nonumber\\
\rho^{\langle g^{2}G^{2}\rangle}&=&-\frac{m_{c}^{2}\langle
g^{2}G^{2}\rangle}{3^{2}\cdot2^{12}\pi^{6}}\int_{\alpha_{min}}^{\alpha_{max}}\frac{d\alpha}{\alpha^{4}}\int_{\beta_{min}}^{1-\alpha}\frac{d\beta}{\beta^{4}}(1-\alpha-\beta)(\alpha^{3}+\beta^{3})\kappa
r^{2},\nonumber\\
\rho^{\langle g\bar{q}\sigma\cdot G q\rangle}&=&\frac{m_{c}\langle
g\bar{q}\sigma\cdot G
q\rangle}{2^{8}\pi^{4}}\Bigg\{-\int_{\alpha_{min}}^{\alpha_{max}}\frac{d\alpha}{\alpha^{2}}\int_{\beta_{min}}^{1-\alpha}\frac{d\beta}{\beta^{2}}(\alpha+\beta-4\alpha\beta)r^{2}
+\int_{\alpha_{min}}^{\alpha_{max}}d\alpha\frac{[m_{c}^{2}-\alpha(1-\alpha)s]^{2}}{\alpha(1-\alpha)}\Bigg\},\nonumber\\
\rho^{\langle\bar{q}q\rangle^{2}}&=&-\frac{m_{c}^{2}\varrho\langle\bar{q}q\rangle^{2}}{3\cdot2^{3}\pi^{2}}\int_{\alpha_{min}}^{\alpha_{max}}d\alpha[m_{c}^{2}-\alpha(1-\alpha)s],\nonumber\\
\rho^{\langle g^{3}G^{3}\rangle}&=&-\frac{\langle
g^{3}G^{3}\rangle}{3^{2}\cdot2^{14}\pi^{6}}\int_{\alpha_{min}}^{\alpha_{max}}\frac{d\alpha}{\alpha^{4}}\int_{\beta_{min}}^{1-\alpha}\frac{d\beta}{\beta^{4}}(1-\alpha-\beta)\kappa
[(\alpha^{3}+\beta^{3})r+4(\alpha^{4}+\beta^{4})m_{c}^{2}]r,\nonumber\\
\rho^{\langle\bar{q}q\rangle\langle g^{2}G^{2}\rangle}&=&\frac{m_{c}\langle\bar{q}q\rangle\langle
g^{2}G^{2}\rangle}{3^{2}\cdot2^{8}\pi^{4}}\int_{\alpha_{min}}^{\alpha_{max}}\frac{d\alpha}{\alpha^{2}}\int_{\beta_{min}}^{1-\alpha}\frac{d\beta}{\beta^{2}}\Big\{(2-\alpha-\beta)(\alpha^{3}+\beta^{3})m_{c}^{2}-3[\alpha^{2}(\beta-1)+\beta^{2}(\alpha-1)]r
\Big\}
,\nonumber
\end{eqnarray}
and
\begin{eqnarray}
\rho^{\langle\bar{q}q\rangle\langle g\bar{q}\sigma\cdot G q\rangle}&=&\frac{m_{c}^{2}\langle\bar{q}q\rangle\langle g\bar{q}\sigma\cdot G q\rangle}{3\cdot2^{5}\pi^{2}}\int_{\alpha_{min}}^{\alpha_{max}}d\alpha(6\alpha^{2}-6\alpha+1),
\nonumber
\end{eqnarray}
where $\kappa=1+\alpha-2\alpha^{2}+\beta+2\alpha\beta-2\beta^{2}$, $r=(\alpha+\beta)m_{c}^2-\alpha\beta s$, $\alpha_{min}=(1-\sqrt{1-4m_{c}^{2}/s})/2$,
$\alpha_{max}=(1+\sqrt{1-4m_{c}^{2}/s})/2$, and $\beta_{min}=\alpha
m_{c}^{2}/(s\alpha-m_{c}^{2})$.
For the four-quark condensate, a general factorization $\langle\bar{q}q\bar{q}q\rangle=\varrho\langle\bar{q}q\rangle^{2}$ \cite{overview1,Narison}
has been employed, in which $\varrho$ may be equal
to 1 or 2.

Equating the two expressions (\ref{Ph}) and (\ref{OPE}), adopting quark-hadron duality, and
making a Borel transform, the sum rule can be turned into
\begin{eqnarray}\label{sumrule}
\lambda^{2}e^{-M_{H}^{2}/M^{2}}&=&\int_{4m_{c}^{2}}^{s_{0}}ds\rho e^{-s/M^{2}}.
\end{eqnarray}
Taking the derivative of Eq. (\ref{sumrule})
with respect to $-\frac{1}{M^{2}}$
and then dividing the result by Eq. (\ref{sumrule}) itself,
one can obtain the hadron's mass sum rule
\begin{eqnarray}\label{sum rule}
M_{H}^{2}&=&\int_{4m_{c}^{2}}^{s_{0}}ds\rho s
e^{-s/M^{2}}/
\int_{4m_{c}^{2}}^{s_{0}}ds\rho e^{-s/M^{2}},
\end{eqnarray}
in which light $u$ and $d$ current-quark masses
have been safely neglected as
they are so small comparing with
the heavy $m_{c}$.

%%%%%%%%%%%%%%%%%%%%%%%%%%%%%%%%%%%%%%%%%%%%%%%%%%%%%%%%%%%%%%%%%%%
\section{Numerical analysis and discussions}\label{sec3}
In the numerical analysis,
the running charm mass $m_{c}$
is $1.27\pm0.02~\mbox{GeV}$ \cite{PDG},
and other input parameters are \cite{svzsum,overview3}:
$\langle\bar{q}q\rangle=-(0.24\pm0.01)^{3}~\mbox{GeV}^{3}$,
$m_{0}^{2}=0.8\pm0.1~\mbox{GeV}^{2}$,
$\langle g\bar{q}\sigma\cdot G q\rangle=m_{0}^{2}~\langle\bar{q}q\rangle$,
$\langle
g^{2}G^{2}\rangle=0.88\pm0.25~\mbox{GeV}^{4}$, as well as $\langle
g^{3}G^{3}\rangle=0.58\pm0.18~\mbox{GeV}^{6}$.
According to the standard criterion
of sum rule analysis, one could
find proper work windows for the threshold parameter $\sqrt{s_{0}}$ and the Borel
parameter $M^{2}$.
The lower bound of $M^{2}$ is obtained from the OPE
convergence, and
the upper one is found in view of that the pole contribution should be larger
than QCD continuum one.
Meanwhile, the threshold
$\sqrt{s_{0}}$ describes the beginning
of continuum state, which is about
$400\sim600~\mbox{MeV}$ bigger than the extracted
$M_{H}$ empirically.

At the very start,
all the input parameters are kept at their central values and the four-quark condensate
factor is taken as $\varrho=1$.
To get the lower bound of $M^{2}$, the OPE
convergence is shown in FIG. 1 by comparing the relative contributions of different
condensates from sum rule (\ref{sumrule}) for $\sqrt{s_{0}}=4.9~\mbox{GeV}$.
Numerically, some main condensates could cancel each other out to some extent
and the relative
contribution of perturbative could play a
predominant role in OPE at $M^{2}=2.5~\mbox{GeV}^{2}$, which is increasing
with the enlarging of Borel parameter $M^{2}$. In this way, it
is taken as $M^{2}\geq2.5~\mbox{GeV}^{2}$
with an eye to the OPE convergence analysis.
Besides, the upper bound of $M^{2}$ is attained with a view to
the pole contribution dominance in phenomenological side.
In FIG. 2, it is compared
between pole contribution and continuum from sum rule (\ref{sumrule})
for $\sqrt{s_{0}}=4.9~\mbox{GeV}$.
The relative pole contribution
is close to $50\%$ at $M^{2}=3.0~\mbox{GeV}^{2}$ and descending with the Borel parameter $M^{2}$.
Thus the pole contribution dominance could be fulfilled
while $M^{2}\leq3.0~\mbox{GeV}^{2}$.
Accordingly, the Borel window of $M^{2}$ is restricted to be
$2.5\sim3.0~\mbox{GeV}^{2}$ for $\sqrt{s_0}=4.9~\mbox{GeV}$.
Analogously, the reasonable window of $M^{2}$ is acquired as $2.5\sim2.9~\mbox{GeV}^{2}$
for $\sqrt{s_0}=4.8~\mbox{GeV}$, and
 $2.5\sim3.2~\mbox{GeV}^{2}$ for $\sqrt{s_0}=5.0~\mbox{GeV}$.
In the work windows, one can expect that the two sides of QCD sum rules have a good
overlap and it is reliable to extract information on the resonance.
The dependence on $M^{2}$ for the mass $M_{H}$ of $Y$ state is shown in FIG. 3,
and its value is computed to be
$4.33\pm0.11~\mbox{GeV}$ in work windows.

\begin{figure}
\centerline{\epsfysize=6.66truecm\epsfbox{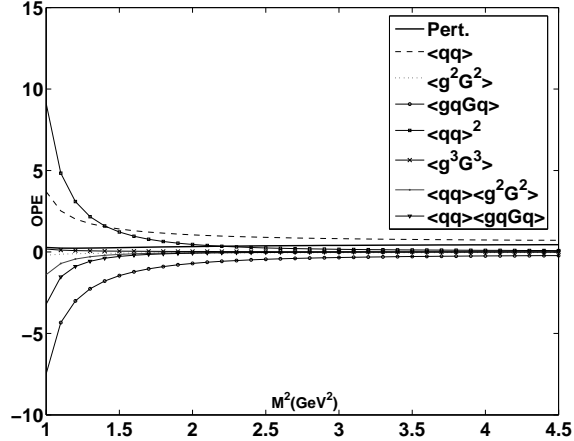}}
\caption{The OPE convergence for the $Y$
state with $P$-wave scalar-scalar $[cq][\bar{c}\bar{q}]$ configuration is shown by comparing the relative contributions of
perturbative, two-quark condensate $\langle\bar{q}q\rangle$, two-gluon condensate $\langle
g^{2}G^{2}\rangle$, mixed condensate $\langle g\bar{q}\sigma\cdot G q\rangle$, four-quark condensate $\langle\bar{q}q\rangle^{2}$, three-gluon
condensate $\langle g^{3}G^{3}\rangle$,
$\langle\bar{q}q\rangle\langle g^{2}G^{2}\rangle$, and
$\langle\bar{q}q\rangle\langle g\bar{q}\sigma\cdot G q\rangle$
from sum rule (\ref{sumrule})
for $\sqrt{s_{0}}=4.9~\mbox{GeV}$.}
\end{figure}

\begin{figure}
\centerline{\epsfysize=6.66truecm\epsfbox{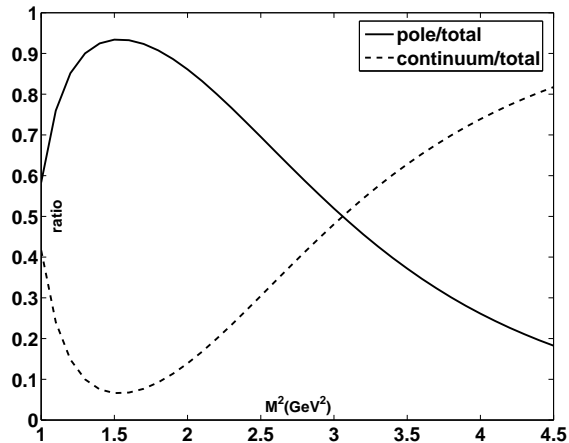}}
\caption{The phenomenological contribution in sum rule
(\ref{sumrule}) for $\sqrt{s_{0}}=4.9~\mbox{GeV}$ for the $Y$
state with $P$-wave scalar-scalar $[cq][\bar{c}\bar{q}]$ configuration.
The solid line is the relative pole contribution (the pole
contribution divided by the total, pole plus continuum contribution)
as a function of $M^2$ and the dashed line is the relative continuum
contribution.}
\end{figure}

\begin{figure}
\centerline{\epsfysize=6.66truecm
\epsfbox{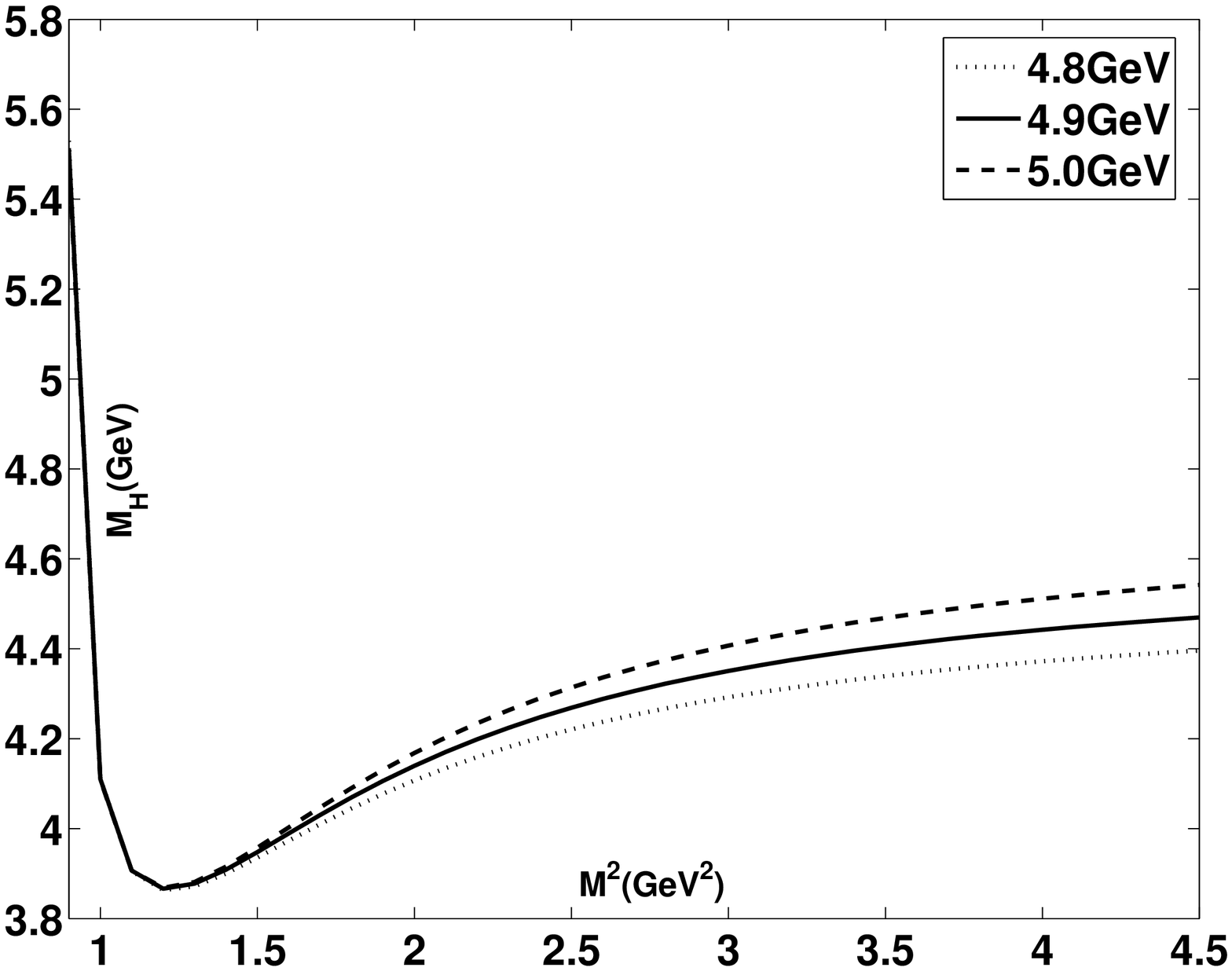}}\caption{The
dependence on $M^2$ for the mass $M_{H}$ of the $Y$
state with $P$-wave scalar-scalar $[cq][\bar{c}\bar{q}]$ configuration
 from sum rule (\ref{sum rule}) is shown while the four-quark condensate
factor $\varrho=1$.
The ranges of $M^{2}$ are $2.5\sim2.9~\mbox{GeV}^{2}$ for
$\sqrt{s_0}=4.8~\mbox{GeV}$, $2.5\sim3.0~\mbox{GeV}^{2}$
for $\sqrt{s_0}=4.9~\mbox{GeV}$, and
$2.5\sim3.2~\mbox{GeV}^{2}$ for $\sqrt{s_0}=5.0~\mbox{GeV}$, respectively.}
\end{figure}

Next varying the input parameters, the mass $M_{H}$ is obtained as
$4.33\pm0.11^{+0.05}_{-0.08}~\mbox{GeV}$ (the first error due
to variation of $s_{0}$ and
$M^{2}$, and the second one resulted from
the uncertainty of QCD parameters) or shortly $4.33^{+0.16}_{-0.19}~\mbox{GeV}$.
In the end, paying attention to the variation of four-quark condensate
factor $\varrho$,
the corresponding Borel
curves are presented in FIG. 4 with $\varrho=2$. In comparison with Fig. 3 for $\varrho=1$,
one could notice the mass uncertainty when varying $\varrho$ from 1 to 2,
and could get the final mass $4.33^{+0.16}_{-0.23}~\mbox{GeV}$ for the $Y$
state with $P$-wave scalar-scalar $[cq][\bar{c}\bar{q}]$ configuration.

\begin{figure}
\centerline{\epsfysize=6.66truecm
\epsfbox{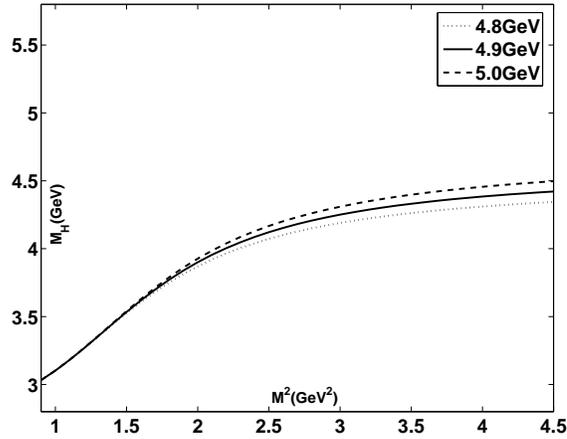}}\caption{The
dependence on $M^2$ for the mass $M_{H}$ of
the $Y$ state with $P$-wave scalar-scalar $[cq][\bar{c}\bar{q}]$ configuration
from sum rule (\ref{sum rule}) is shown while the four-quark condensate
factor $\varrho=2$.}
\end{figure}

In experiment, one may note that in the hidden-charm
$e^{+}e^{-}\rightarrow\gamma_{ISR}\pi^{+}\pi^{-}\psi(2S)$ process,
BABAR observed a broad structure near $4.32~\mbox{GeV}$ \cite{Y4360-1}, and
Belle subsequently found the charmonium-like state $Y(4360)$ \cite{Y4360-2}.
Afterward, a combined fit to these cross
sections measured by BABAR and Belle experiments was performed \cite{Y4360-fit},
and the property of $Y(4360)$ was further studied in $e^{+}e^{-}\rightarrow\pi^{+}\pi^{-}\psi(2S)$
via initial-state radiation at BABAR \cite{Y4360-3} and at Belle \cite{Y4360-4}.
Taking notice of the close masses of $Y(4360)$ and $Y$ state concerned here,
one could conjecture that they may be the same structure attributing to different
decay modes.
If that true, it would be very important for understanding $Y(4360)$
to search for the predicted $Y$ state,
because complementary measurements
by other decay modes such as the open-charm process will provide further
insights into $Y(4360)$'s internal structure.
Whether or not, it is undoubtedly exciting and significative if
one could find a vector charmonium-like $Y$
state particularly in an open-charm decay.

Invigoratingly, there has appeared some measurement of
Born cross section for $e^{+}e^{-}\rightarrow D^{-}D_{1}(2420)^{+}+c.c.$ \cite{DD-Ex},
in which the cross section line shape is consistent
with the previous BESIII's result based on full reconstruction method \cite{DD-Ex1},
and there is some indication of
enhanced cross section at the location of $Y(4360)$.
Thereby, it seems promising that the predicted $Y$
state could be observed in the open-charm process $e^{+}e^{-}\rightarrow\bar{D}D_{1}(2420)+c.c.$
via either the initial-state radiation or the direct production for the future experiments.

%%%%%%%%%%%%%%%%%%%%%%%%%%%%%%%%%%%%%%%%%%%%%%%%%%%%%%%%%%%%%%%%%%%
\section{Summary}\label{sec4}

Activated by the first observation of a vector
charmonium-like state $Y(4626)$ in the open-charm $D_{s}^{+}D_{s1}(2536)^{-}$ decay mode,
for which could be a $P$-wave scalar-scalar $[cs][\bar{c}\bar{s}]$ tetraquark state,
we predict a novel vector
charmonium-like $Y$ state with $P$-wave scalar-scalar $[cq][\bar{c}\bar{q}]$ configuration.
Finally, the mass of $Y$
is presented to be $4.33^{+0.16}_{-0.23}~\mbox{GeV}$
from QCD sum rules.
We suggest that the predicted $Y$ state
could be searched for in an open-charm $e^{+}e^{-}\rightarrow\bar{D}D_{1}(2420)+c.c.$ process
through the initial-state radiation or the direct production in experiments,
for which virtually there has been some indication of
enhanced cross section in BESIII's existing measurements.
%%%%%%%%%%%%%%%%%%%%%%%%%%%%%%%%%%%%%%
\begin{acknowledgments}
This work was supported by the National
Natural Science Foundation of China under Contract
Nos. 11475258 and 11675263, and by the project for excellent youth talents in
NUDT.
\end{acknowledgments}

%%%%%%%%%%%%%%%%%%%%%%%%%%%%%%%%%%%%%%%%%%%%%%%%%%%%

\end{document}